\documentclass[aps,prb,superscriptaddress,showkeys,longbibliography,twocolumn]{revtex4-2}

\usepackage{graphicx, amsmath, amssymb, bm, upgreek, amsfonts, bbm} 
\usepackage[colorlinks=true, linkcolor=blue, citecolor=blue,          urlcolor=blue]{hyperref}
\usepackage{xcolor}
\usepackage{booktabs}
\def\vp{{\mathbf p}}
\def\vr{{\mathbf r}}
\def\va{{\mathbf a}}
\def\vk{{\mathbf k}}
\def\vu{{\mathbf u}}

\def\vv{{\mathbf v}}

\def\vd{\delta \vp}
\def\mj{{\mathbbm{j}}}
\def\mJ{{\mathbb J}}

\def\mD{{\mathbb D}}
\def\mA{{\mathbb A}}
\def\mR{{\mathbb R}}

\begin{document}

\title{Exact mean-field phase diagram for self-avoiding active particles in a lattice}
\author{Felipe Hawthorne}
\affiliation{Department of Physics, Federal University of Parana,
               R. Evaristo F. Ferreira da Costa, 81530-015, Curitiba, Brazil}
\affiliation{Interdisciplinary Center for Science, Technology, and Innovation (CICTI), Federal University of Parana, Av. Cel. Francisco H. dos Santos, 81530-000, Curitiba, Brazil}

\author{Cristiano F. Woellner}
\affiliation{Department of Physics, Federal University of Parana,
               R. Evaristo F. Ferreira da Costa, 81530-015, Curitiba, Brazil}
\affiliation{Interdisciplinary Center for Science, Technology, and Innovation (CICTI), Federal University of Parana, Av. Cel. Francisco H. dos Santos, 81530-000, Curitiba, Brazil}

\author{Jos\'e A. Freire}
\email[]{jfreire@fisica.ufpr.br}
\affiliation{Department of Physics, Federal University of Parana,
               R. Evaristo F. Ferreira da Costa, 81530-015, Curitiba, Brazil}

\begin{abstract}
We investigate motility-induced phase separation in a lattice gas of self-propelled particles with hard-core exclusion, where an internal director biases particle hopping along the lattice coordination directions while undergoing rotational diffusion, together with a thermal-like translational diffusion. Rather than employing stochastic simulations, we adopt a master-equation formalism within a general mean-field approximation. By linearizing the mean-field master equation around the homogeneous stationary state and applying Bloch's theorem, the stability analysis is reduced to a $z$-dimensional tight-binding eigenvalue problem. A perturbation expansion in the wavenumber near $\vk = 0$ then yields the spinodal surface in closed analytical form for six Bravais lattices: linear, square, hexagonal, simple cubic, body-centered cubic, and face-centered cubic. The influence of lattice geometry is shown to enter exclusively through a single coefficient $\mathcal{A}$ which we evaluate exactly for each case. We further show that translational diffusion smooths the interface between the dense and dilute phases. Finally, we determine the rotational probability currents associated with the inhomogeneous stationary states, a distinctive signature of the broken detailed balance underlying active-system dynamics.
\end{abstract}

\maketitle
\section{Introduction}
Self-propelled particles with purely repulsive interactions can phase-separate into coexisting dense and dilute states via a kinematic mechanism independent of attraction or alignment: particles accumulate where they slow down, increasing local density and further reducing their speed. This instability was first identified by Tailleur and Cates in a one-dimensional run-and-tumble model with density-dependent tumbling rates~\cite{Tailleur_2008}. Simulations of persistently moving and repulsively interacting particles later confirmed the same behavior~\cite{Fily_2012,Redner_2013}, now known as Motility-Induced Phase Separation (MIPS). Experimental evidence was reported in active colloids~\cite{Buttinoni_2013}, with comprehensive reviews in~\cite{Marchetti_2013,Cates_2015,Gompper_2020}.

MIPS on lattices have been studied in settings that include  density-dependent tumbling rates~\cite{Thompson_2011,Solon_2013,Solon_2015}, and persistent motion plus hard-core exclusion~\cite{Soto_2014,Dittrich_2021,Whitelam_2018,Yao_2025}.

Although the bulk of MIPS theory has been developed through Langevin-type simulations and effective hydrodynamic field equations~\cite{Cates_2013,Stenhammar_2013}, the complementary probabilistic approach, based on master equations or Fokker-Planck equations, has remained comparatively rare in the MIPS literature, see however~\cite{Bialke2013}. 

We consider the master equation for self-propelled particles on a generic lattice with hard-core exclusion. Each particle carries an internal director that biases its persistent hopping along one of the $z$ lattice directions and undergoes rotational diffusion, alongside thermal-like translational diffusion.

We present in close analytical form, for six different lattices and at the mean-field level, the spinodal surface that encloses the region  of parameter space where the homogeneous state (gas phase) becomes locally unstable. 

The probabilistic description of MIPS provides a complementary perspective on clustering, distinct from the dynamical, Langevin-based approaches prevalent in the literature. Within this framework, the roles of active motion and thermal-like noise are captured explicitly and visualized directly, making their contributions to phase separation transparent.

A hallmark of active matter is the presence of a nonzero, purely rotational, steady-state probability current in configuration space, arising from the breakdown of detailed balance~\cite{Cates_2012,Fang_2019,Fang_2020,Fodor_2016,Nardini_2017} and observed in biological systems~\cite{Angelani_2009,Di_Leonardo_2010,Battle_2016}. Such currents are not directly accessible within Langevin or hydrodynamic descriptions, whereas they emerge here transparently.

This paper is organized as follows. Section \ref{sec2} defines the lattice model. Section \ref{sec3} derives the mean-field master equation, obtains the closed-form spinodal surface for six Bravais lattices, and examines the inhomogeneous stationary states and their rotational probability currents. Section \ref{sec4} closes with a summary and outlook.

\section{The Active Lattice Gas Model}
\label{sec2}
In this section we define the model in terms of its stochastic dynamics, and in the next section the corresponding master equation. Our work will be entirely focused on the later.

We consider $N$ active particles moving by hopping between nearest-neighbor sites of a Bravais lattice (linear, square, hexagonal, simple cubic, body-centered cubic and face-centered cubic) . Steric exclusion is enforced: no two particles may occupy the same site simultaneously. Each particle carries an internal director that, at any given time, points toward one of the $z$ (the lattice coordination number) directions.

Our model closely follows the one introduced in Ref.~\cite{Dittrich_2021}. Both the particle positions and their internal directors evolve in discrete time steps via the following update rules:
\begin{enumerate}
\item Active motion: a particle attempts to move in the direction of its current internal director with rate $w_a$. That is, if the target site is unoccupied, the hop occurs with probability $w_a\,dt$, where $dt$ is an infinitesimal time increment.

\item Translational diffusion: independently of its internal director, a particle attempts to move to a randomly chosen nearest-neighbor site with rate $w_t$. As before, the move is accepted only if the target site is empty.

\item Rotational diffusion of the director: the internal director can rotate to any of its adjacent directions with rate $w_r$. Our adjacency criterium is of minimal angle between lattice directions. For instance, on a square lattice, if the current direction is ``up", the director may rotate to ``left" or ``right" with probability $w_r\,dt$ each.
\end{enumerate}

\section{The Mean-Field Master Equation}
\label{sec3}

At the mean-field level, the probability of observing a given global microstate at time \( t \) is expressed in terms of the local probabilities \( p_{\vr,s}(t) \), which represent the likelihood that site \( \vr \) is occupied by a particle whose director points in the \( s \)-th direction, with \( s = \{1, \ldots, z \}\).

It follows that, 
\begin{equation} 
p_{\vr,\Sigma}\equiv \sum_{s=1}^z p_{\vr,s}
\end{equation}
is the probability that site $\vr$ is occupied and 
\begin{equation}
    N=\sum_{\vr} p_{\vr,\Sigma} \label{N}
\end{equation}
is the total number of particles.

The time evolution of these probabilities follows from the stochastic dynamics defined above.
\begin{equation}
\begin{aligned}\label{ME}
  \frac{dp_{\vr,s}}{dt} = & ~ w_a \, \big[ p_{\vr-\va_s, s} \, (1 - p_{\vr,\Sigma}) - p_{\vr, s} \, (1 - p_{\vr+\va_s,\Sigma}) \big] \\
+ & ~ w_t \,\sum_{\vr' \sim \, \vr} \big[ p_{\vr', s} \, (1 - p_{\vr,\Sigma})
-  p_{\vr,s} \, (1 - p_{\vr', \Sigma} ) \big] \\
+ & ~ w_r \, \sum_{s' \sim \,s} (p_{\vr,s'} - p_{\vr,s}).
\end{aligned}
\end{equation}

Here, $\va_s$ denotes the lattice vector in the coordinate direction $s$. The notation $\vr' \sim \, \vr$ indicates that $\vr'$ is a nearest neighbor to $\vr$, and $s' \sim s$ indicates that $s'$ is one of the coordinate directions adjacent to $s$ according to a minimal angle criterion.

The first term on the right-hand side represents active motion along the director $\va_s$; the second term accounts for translational diffusion; and the third term describes the rotational diffusion of the director. The $(1-p)$ factors enforce the steric interaction.

This set of equations conserves the total number of particles and has a trivial homogeneous stationary solution (for a general lattice with $M$ sites and coordination number $z$)
\begin{equation}
    p^H_{\vr,s} = \phi/z, \label{H solution}
\end{equation}
where \( \phi \) is the filling fraction
\begin{equation}
    \phi\equiv N/M.
\end{equation}

\subsection{The local stability of the homogeneous solution}
Here we obtain analytically the domain of parameter space $\{w_a, w_t, \phi\}$ ($w_a$ and $w_t$ in units of $w_r$) where the homogeneous stationary solution is locally stable. 

Linearizing the master equation about the homogeneous solution, \( \vp(t) = \vp^H + \vd(t) \), where the $Mz$ probabilies are arranged as a single-indexed vector with the $z$ internal degrees of freedom grouped together for each of the $M$ lattice sites, i.e., $\vp=(p_{11},\ldots,p_{1z}, \ldots, p_{M1},\ldots, p_{Mz})$, 
\begin{equation}
    \frac{d\vd}{dt} = \mJ^H \cdot \vd.
\end{equation}

The Jacobian matrix of the homogeneous solution, denoted \( \mJ^H \), consists of \( z \times z \) blocks \( \mJ^H_{\vr,\tilde \vr} \) that couple internal degrees of freedom at neighboring lattice sites \( \vr \) and \( \vr' \). More precisely:
\begin{equation}
    \mJ^H_{\vr,\tilde \vr} = 
    \begin{cases}
        -w_t z\, \mD + w_r\,\mR - w_a \sum\limits_{s=1}^z \mA(s), & \text{if } \tilde \vr = \vr, \\
        w_t\, \mD + w_a\, \mA(s), & \text{if } \tilde \vr = \vr + \va_{s},
    \end{cases}
\end{equation}
where the $z$-dimensional square matrices \( \mD \), \( \mR \), and \( \mA(s_0) \) are:
\begin{equation}
    \mD_{s,\tilde s} = (1-\phi)\,\delta_{s,\tilde s}  + \tfrac{\phi}{z}
\end{equation}

\begin{equation}
    \mR_{s,\tilde s} = -n_z \, \delta_{s,\tilde s} + \delta_{s\sim\tilde s}
\end{equation}

\begin{equation}
    [\mA(s_0)]_{s,\tilde s} = (1-\phi)\,\delta_{s,-s_0}\, \delta_{s,\tilde s} + \tfrac{\phi}{z}\, \delta_{s,s_0}
\end{equation}

The $n_{z}$ in the $\mR$ matrix is the number of adjacent lattice coordinate directions (e.g., $n_z=2$ in the square and hexagonal lattices). $\delta_{s,\tilde s}$ is the usual Kronecker delta, and $\delta_{s\sim\tilde s}$ is $1$ if $s$ and $\tilde s$ are adjacent directions, and $0$ otherwise. The matrix $\mR$ can be viewed as the Laplacian of the (undirected) connection graph of coordinate directions.

The eigenvalues of the Jacobian matrix determine the local stability of the homogeneous stationary state \eqref{H solution}. The translational invariance of $\mJ^H$ (assuming periodic boundary conditions (PBC)) implies that its eigenvectors may be searched in Bloch form
\begin{equation}
    (\delta\vp_\vk)_{\vr,s} = (\vu_\vk)_s\, e^{i\vk\cdot\vr},
\end{equation}
where the $M$ vectors $\vk$ (consistent with the PBC) are restricted to the first Brillouin zone (1BZ) of the Bravais lattice.
The $z$ eigenvalues associated with each $\vk$ are obtained from the following $z$-dimensional eigenvalue problem:
\begin{equation}
\begin{aligned}
    \mj_\vk \cdot \vu_\vk = & \, \lambda_\vk \,\vu_\vk \\
    \mj_\vk \equiv \Big\{ w_r \, \mR + w_t \,F(\vk) \,\mD \, +& \, w_a \sum_s (e^{i\vk\cdot \va_s}-1)\,\mA(s) \Big\} \label{H}\\
    F(\vk) \equiv & \sum_s (e^{i\vk\cdot \va_s} - 1)
\end{aligned}
\end{equation}

The entire procedure is analogous to analyzing the spectrum of a tight-binding Hamiltonian with \( z \) orbitals per lattice site. In particular, the eigenvalue spectrum is invariant under lattice point symmetries,
\begin{equation}
\lambda_{S\vk} = \lambda_{\vk}, \label{S}
\end{equation}
where \( S \) denotes a matrix representation of any of the elements of the lattice point symmetry group.

$\mj_{\vk=0} = w_r \, \mR$ and exhibits a trivial zero eigenvalue associated with a uniform (all-ones) eigenvector. For a generic wavevector $\vk$, some of the eigenvalues of $\mj_\vk$ may be complex. The homogeneous stationary state is locally unstable when any of the $Mz$ eigenvalues of the full matrix $\mJ^H$ has a positive real part. 

The full spectrum is naturally presented in $\vk$ space and is illustrated in the Supporting Information for the case of a square lattice. There it becomes apparent that the negative real part eigenvalues first appear in the immediate vicinity of $\vk=0$. That is, the homogeneous state becomes unstable against large wave-length fluctuations.

\subsection{The spectrum near $\vk=0$}

Here we assume an infinite lattice, so that the wavevectors $\vk$ form a continuum within the first Brillouin zone.

The matrix $\mj_\vk$ can be expanded in powers of $k$ near $\vk=0$:
\begin{equation}
\mj_{\vk \sim 0} = \, w_r \, \mR + \mj_1 + \mj_2 + \ldots
\end{equation}
where ($d$ is the spatial dimension and $a=|\va_s|$ is the lattice parameter)
\begin{equation}
\begin{aligned}
    \mj_1 \equiv & \, w_a \, \sum_s i(\vk\cdot \va_s)\, \mA(s) \\
    \mj_2 \equiv & -  \frac{w_t z a^2 \vk^2}{2d} \, \mD -  \,
        \frac{w_a}{2} \sum_s (\vk\cdot \va_s)^2\, \mA(s)
\end{aligned}
\end{equation}

Standard Rayleigh–Schrödinger perturbation theory, applied with the caveat that $\mj_1$ and $\mj_2$ are non-Hermitian matrices, provides the leading-order $k$-dependence of the zero eigenvalue.

In terms of the normalized eigenvectors of the $\mR$ matrix, 
\begin{equation}
    \mR \cdot \vv_n  = \lambda_n \vv_n, \quad (n=0,...,z-1),
\end{equation}
where, in particular,
\begin{equation}
    \vv_0 = z^{-1/2} [1,1,\ldots,1]^T \qquad (\lambda_0=0),
\end{equation}
we obtain near $\vk = 0$:
\begin{equation}
\begin{split}
    \lambda_{0} &= \vv_0^\dagger \cdot \mj_1 \cdot \vv_0 + \vv_0^\dagger \cdot \mj_2 \cdot \vv_0\\ - &\sum_{n\ne 0} \frac{(\vv_0^\dagger \cdot \mj_1 \cdot \vv_n)  (\vv_n^\dagger \cdot \mj_1 \cdot \vv_0)}{w_r \lambda_n}+\mathcal{O}(k^3) \label{PertTheory1}
    \end{split}
\end{equation}
In the Supporting Information we computed \eqref{PertTheory1} for six different Bravais lattices and found:
\begin{equation}\label{A}
\begin{split}
\lambda_0 &= \Bigg[ -\bigg( \frac{w_t z + w_a}{2d}\bigg) + \frac{w_a^2(1-\phi)(1-2\phi)}{w_r\,z}\,{\cal{A}}\Bigg]\,\vk^2,\qquad \\&(\vk \sim 0).
\end{split}
\end{equation}
The values of ${\cal{A}}$ for the six Bravais lattices here considered are listed in Table \ref{tab:param}.

\begin{table}[t]
\centering
\begin{tabular}{cc}
\toprule
Lattice (space dimension) & ${\cal{A}}$ \\
\midrule
linear (1d)  & $-1$ \\
square (2d)  & $-1$ \\
hexagonal (2d) & $-3$ \\
simple cubic (3d) & $-1/2$ \\
body-centered cubic (3d) & $-4/3$ \\
face-centered cubic (3d) & $-2$\\
\bottomrule
\end{tabular}
\caption{Values of ${\cal{A}}$ in eq. \eqref{A} for some Bravais lattices}\label{tab:param}
\end{table}

Owing to eq. \eqref{S}, the spectrum near $\vk = 0$ was expected to be spherically symmetric for all Bravais lattices in Table \ref{tab:param}.

\begin{figure}[h]
\centering
{\includegraphics[width=0.45\textwidth]{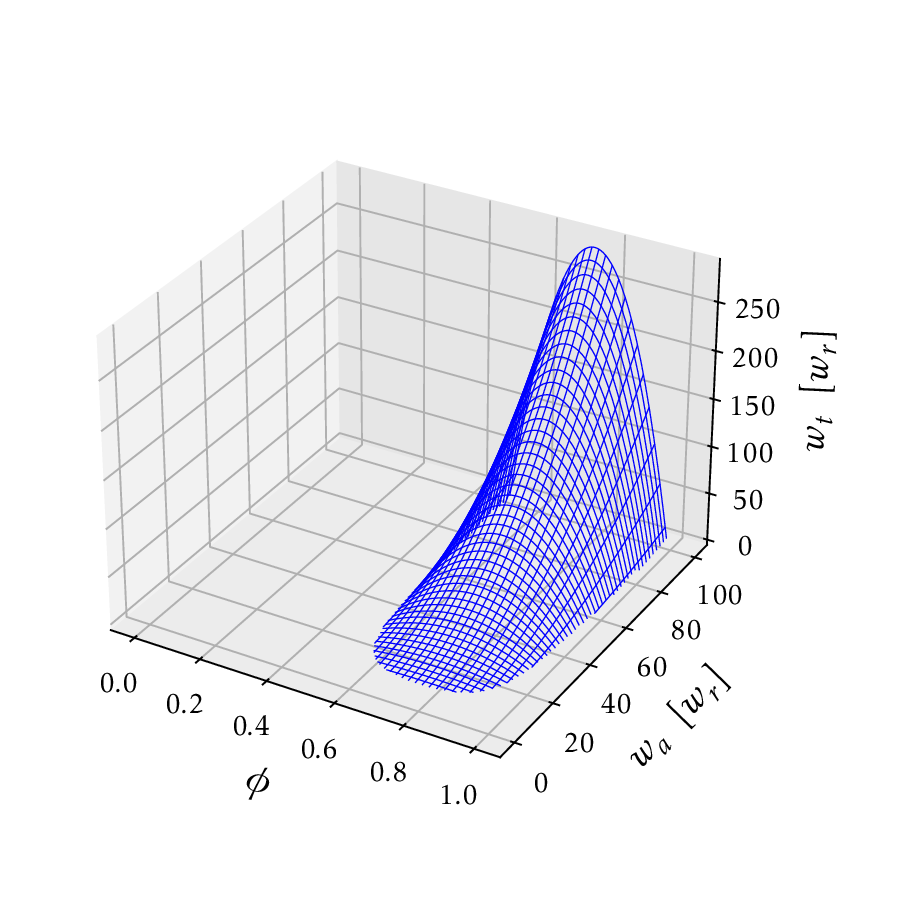}}
\caption{The spinodal surface in parameter space for the square lattice. $w_a$ and $w_t$ are the active motion and translational diffusion rates respectively (both expressed in terms of the rotational diffusion rate $w_r$), $\phi$ is the lattice filling fraction. The homogeneous stationary state is locally unstable inside the spinodal surface. Examples of locally stable stationary states inside the surface are shown in Fig. \ref{Fig:2}.} \label{Fig:1}
\end{figure}

The region in parameter space $\{w_a/w_r, w_t/w_r, \phi\}$ where the bracket term in eq. \eqref{A} vanishes is the mean-field version of the spinodal surface. Inside the spinodal surface some of the eigenvalues of $\mJ^H$ has positive real part and the homogeneous state is locally unstable. Figure \ref{Fig:1} illustrates the case of the (infinite) square lattice. 

In Ref.~\cite{Kourbane_Houssene_2018}, an analytical expression for the spinodal surface is derived, in the hydrodynamic limit, for a one-dimensional model closely related to ours. By contrast, our result is exact at the mean-field level and explicitly incorporates the structure of the underlying Bravais lattice through the parameter ${\cal A}$.

For finite lattice systems, the wavevectors $\vk$ form a discrete set within the 1BZ and the stability analysis has to proceed numerically. Typically, the spinodal surface is slightly ``contracted" compared to the infinite-lattice result, implying that, for fixed $w_t/w_r$ and $\phi$, larger values of $w_a/w_r$ are needed to destabilize the homogeneous state.

\subsection{The Spinodal region}

\begin{figure}[h!]
\centering
{\includegraphics[width=0.6\columnwidth]{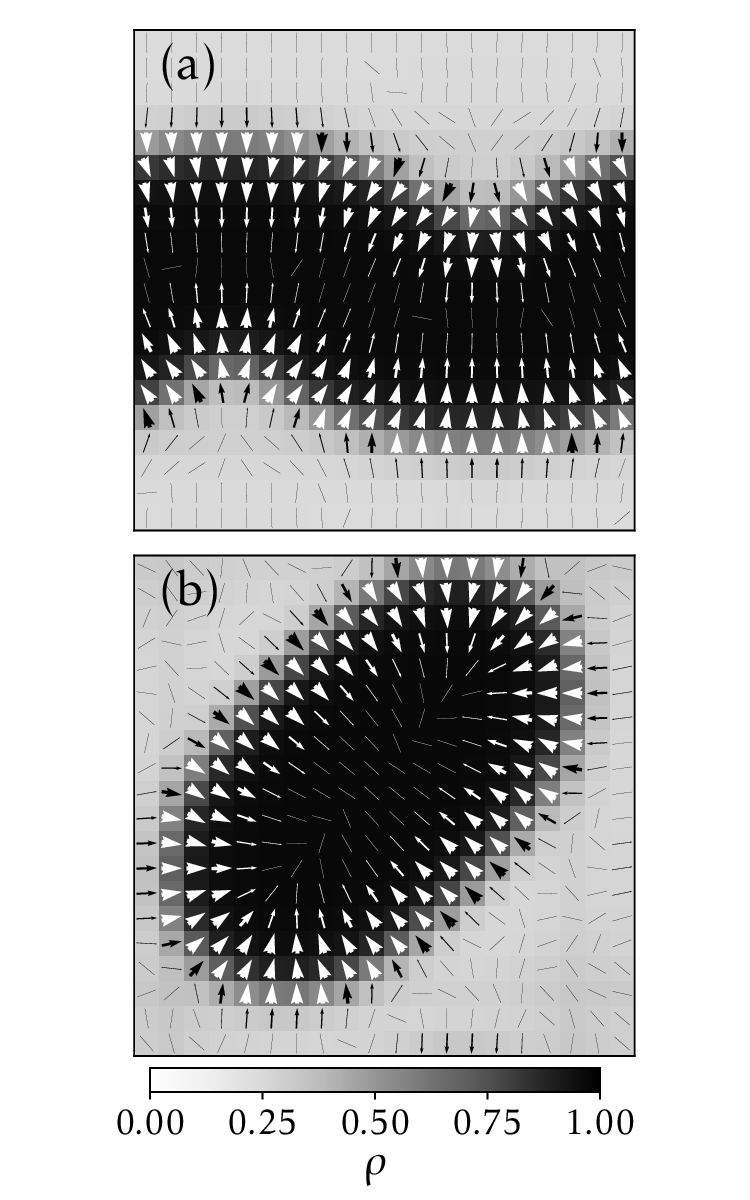}}
\caption{Average site occupation (grayscale) and persistence velocity (arrow thickness) for two stationary solutions of the ME at ${w_t=0, w_a=20\,w_r, \phi=0.6}$ on a $20\times20$ square lattice. The homogeneous state is locally unstable for these parameters, while both solutions are locally stable and represent distinct manifestations of motility-induced phase separation. Persistent motion effectively “compresses” the cluster.}\label{Fig:2}
\end{figure}

Figure \ref{Fig:2} illustrates two different inhomogeneous stationary solutions of the mean field master equation for a $20\times 20$ square lattice, both obtained with the same set of model parameters where the homogeneous solution is unstable (checked numerically), i.e., within the spinodal region. We simply integrated Eq.\eqref{ME} until convergence, starting from two different initial conditions. The average site occupation, $\sum_s p_{\vr,s}$ in shown as gray tones and the average persistence velocity, $\sum_s \hat{\va}_s\, p_{\vr,s}$, as arrows. The persistent motion is concentrated at the cluster boundary, and directed towards the cluster, as expected.

These solutions are locally stable, i.e., the numerically computed Jacobian spectrum consists of a single zero eigenvalue, with all other eigenvalues having negative real parts.

Because the master equation is invariant under the lattice space group operations, any stationary state is mapped to another stationary state by an arbitrary lattice translation combined with any lattice point-group symmetry. Each such stationary solution possesses its own basin of attraction and can be reached from suitable initial conditions.

\subsubsection{The effect of the translational diffusion}
\begin{figure*}[tp]
\centering
{\includegraphics[width=0.65\textwidth]{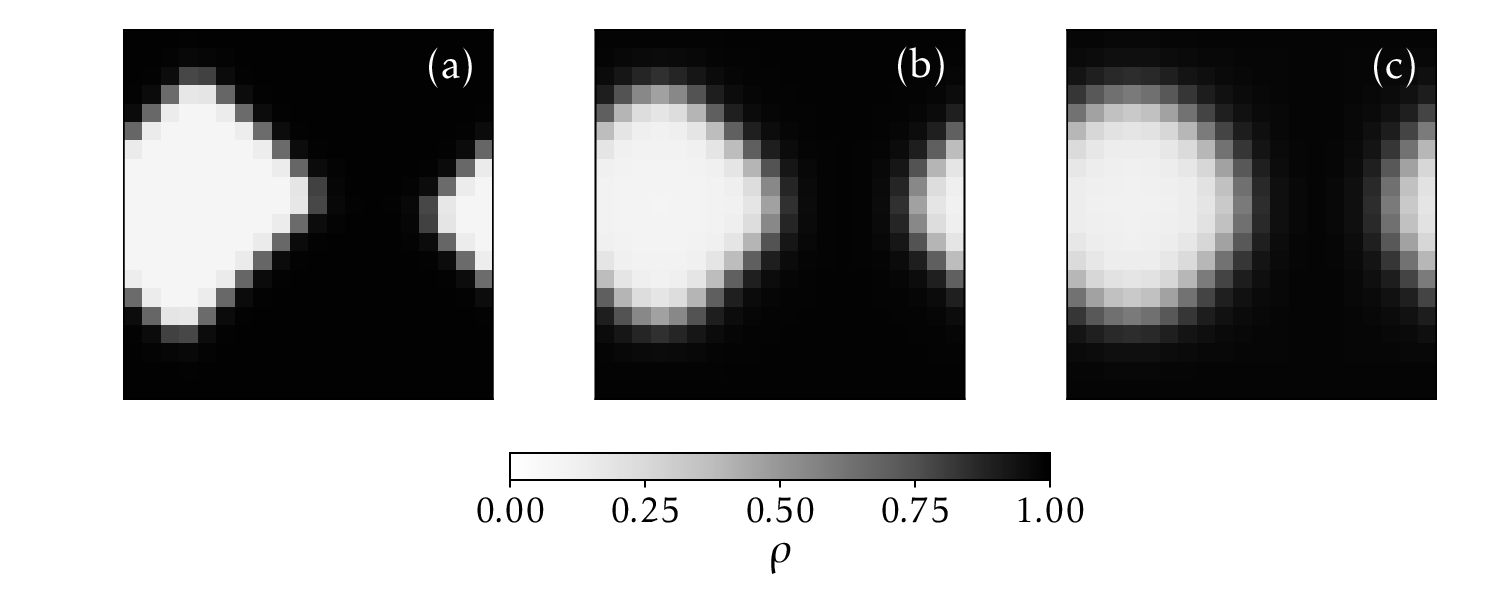}}
\caption {Average site occupation (grayscale) for three stationary ME solutions in a $20\times 20$ square lattice at $w_a=60,w_r$, $\phi=0.75$, and increasing $w_t$: (a) $0$, (b) $5\,w_r$, (c) $10\,w_r$. Motility-induced phase separation appears as a gas pocket within a dense phase; as $w_t$ increases, the interface becomes progressively more diffuse.} \label{Fig:3}
\end{figure*}

Still in a $20\times 20$ square lattice, Figure \ref{Fig:3} shows the average site occupation of three locally stable stationary solutions of the ME, with parameters $w_a=60\,w_r, \phi=0.75$ and $w_t=0$ in panel (a), $w_t=5\,w_r$ in panel (b), and $w_t=10\,w_r$ in panel (c). All three parameters deep inside the domain of instability of the homogeneous state. 

The figure shows the softening effect of $w_t$ on the interface between the gas and the condensed phases, something that was also observed and discussed in the context of hard disks in a Langevin type simulation~\cite{Hawthorne_2025}. 

For large enough $w_t$, see Fig. \ref{Fig:1}, one crosses vertically the spinodal surface and the homogeneous state becomes locally stable again. 

Whereas active motion ($w_a$) promotes MIPS, translational diffusion ($w_t$) hinders it.

\subsubsection{The rotational probability current}
The master equation can be viewed as a discrete probability continuity equation, and its stationary states have divergenceless probability currents (see the lattice expression for the current in the SI).  All inhomogeneous stationary solutions of the ME display a non-zero, hence purely rotational, probability current. This is to be contrasted with relaxation to equilibrium when the master equation obeys detailed balance (not the case with Eq.~\eqref{ME}). When detailed balance is present, the ME drives the system to free-energy minima, with zero probability current. See the theoretical discussion in Refs.~\cite{Fang_2019, Fang_2020}, and an experimental illustration in Ref.~\cite{Battle_2016}.

\begin{figure}[h!]
\centering
{\includegraphics[width=0.6\columnwidth]{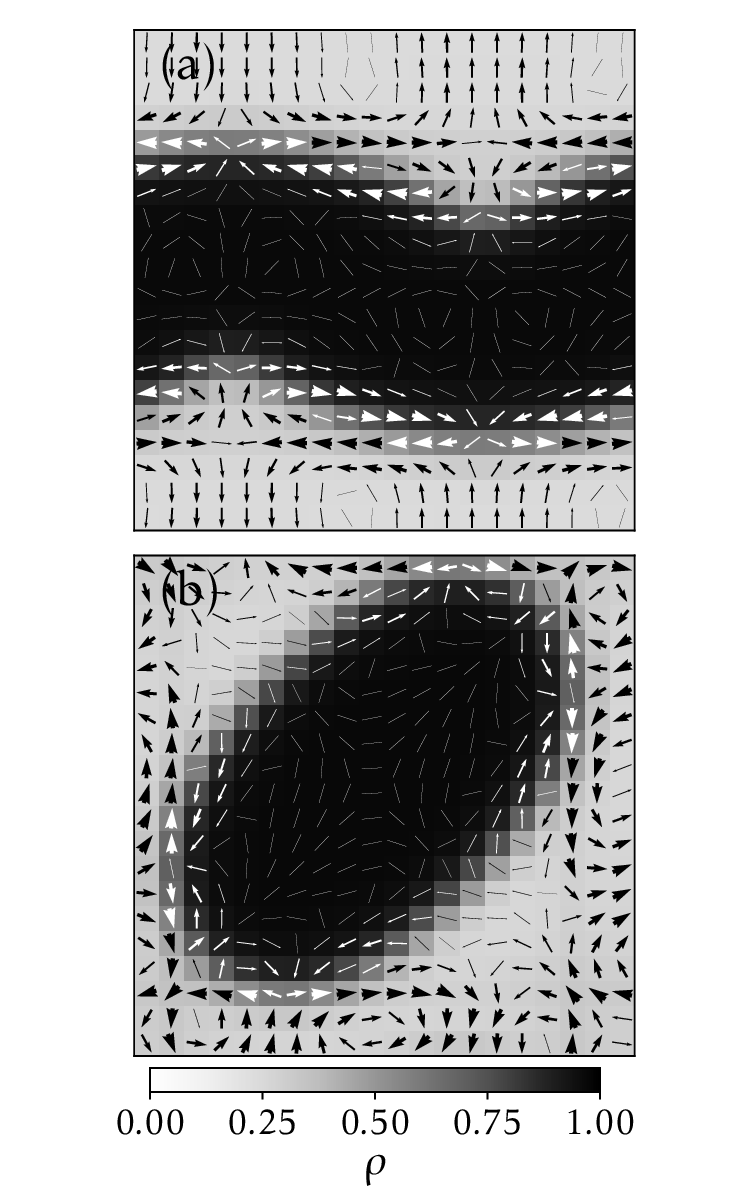}}
\caption{Average site occupation (grayscale) and probability current (arrow thickness) of the two inhomogenous stationary states shown in Fig.~\ref{Fig:2}. The rotational steady probability current in the low concentration region.}\label{Fig:4}
\end{figure}

Figure \ref{Fig:4} illustrates the probability current of the two inhomogeneous stationary states in Fig.~\ref{Fig:2}. The concentration of the probability current in the gas phase region is the probabilistic manifestation of the continuous attachment/detachment of particles to the MIPS cluster commonly seen in Langevin type simulations, see for instance Ref.~\cite{Hawthorne_2025}.    

\subsubsection{Homogeneous state escape time}
We consider the mean escape time of an initial condition in the vicinity of the unstable homogeneous steady state. Following this escape, trajectories evolve toward a stable inhomogeneous state. The escape time is governed by the number and magnitude of the eigenvalues of $\mJ^H$ with positive real parts, as well as by the projection of the initial condition onto the corresponding unstable eigenspace. We adopt a specific average measure of this timescale and analyze its dependence on model parameters within the spinodal region. 

We use as initial condition:
\begin{equation}
    p_{\vr,s}(0) = \tfrac{\phi}{z} + \xi_{\vr,s},
\end{equation}
with $\xi_{\vr,s}$ uniformly distributed with zero average and $\sum_{\vr,s} \xi_{\vr,s}^2 = \epsilon^2$. That is, the starting point is at an Euclidean distance of $\epsilon$ from the homogeneous steady state. 

We followed the time evolution of the decrease of the dimensionless entropy per particle, 
\begin{equation}
    \Delta S(t) \equiv - \frac{1}{N} \sum_{\vr,s} p_{\vr,s}(t) \log [p_{\vr,s}(t)] - S_H,
\end{equation}
where $S_H=-\log (\phi/z)$ is the homogeneous state entropy per particle. Some representative entropy evolutions can be seen in the SI.

\begin{figure*}[tp]
\centering
{\includegraphics[width=0.7\textwidth]{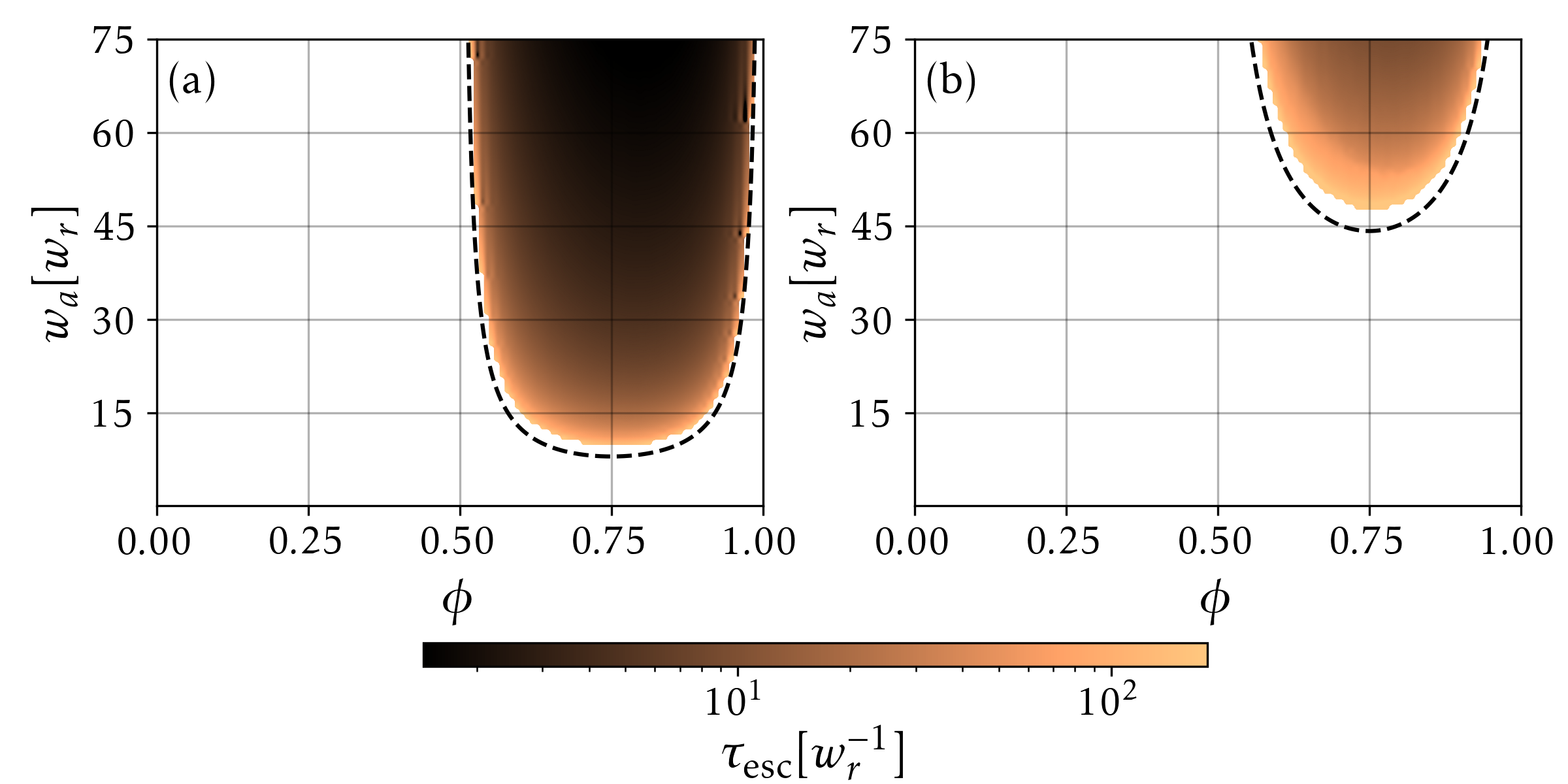}}
\caption{The average escape time from the locally unstable homogeneous state in a $80\times 80$ square lattice. Panel (a) corresponds to $w_t=0$, and panel (b) to $w_t=50\,w_r$. The dashed lines are the infinite square lattice spinodal lines at the corresponding value of $w_t$. As expected, the homogeneous state becomes increasingly unstable, reflected in a decreasing $\tau_{esc}$, as one moves deeper into the spinodal region.}\label{Fig:5}
\end{figure*}

We performed an average over many realizations of the departing point (standard deviations reported in the SI) and defined the average escape time as the time when the first inflexion point of $\Delta S(t)$ is attained. The results, obtained for different model parameters, are shown in Fig. \ref{Fig:5}.

\subsection{The Metastable Region}

The metastable region is the portion of parameter space outside the spinodal surface and inside the binodal surface in which the homogeneous state remains linearly stable yet coexists with locally stable inhomogeneous stationary states. If an underlying free-energy function existed, this region would be characterized by multiple free-energy minima. Outside the binodal surface the ME has a single locally stable stationary state, the homogeneous state.

\begin{figure}[h]
\centering
{\includegraphics[width=0.6\columnwidth]{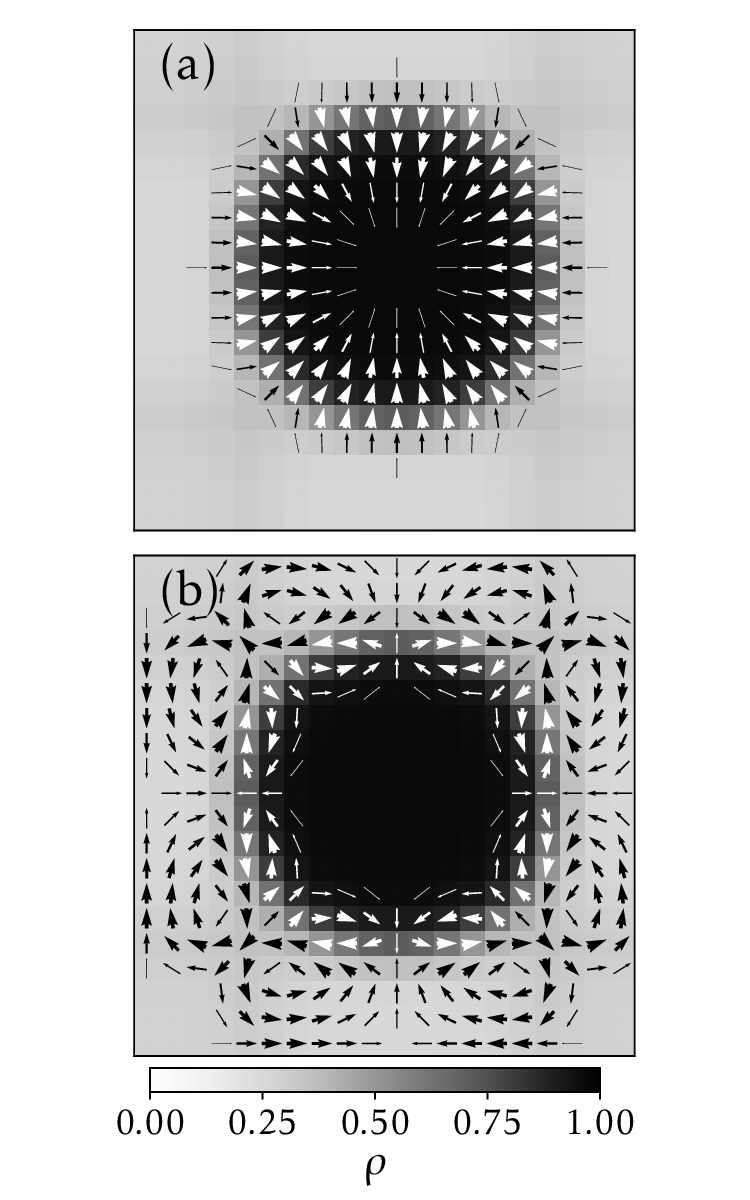}}
\caption{Average site occupation (gray tones) and (a) average site persistence velocity; and (b) site probability current (in both cases, magnitude indicated by arrow thickness) of a particular, locally stable, inhomogeneous stationary state of the ME obtained with $\{w_t=0, w_a=20\,w_r,\phi=0.5\}$, i.e., outside the spinodal surface. For this set of parameters, the homogeneous state is also locally stable. This solution shares the same MIPS features of the ones shown in Figs. \ref{Fig:2} and \ref{Fig:4}.}\label{Fig:6}
\end{figure}

A systematic investigation of the $Mz$ dimensional microstate space to locate the binodal surface is beyond reach, so that we simply illustrate one such inhomogeneous stationary state for a set of parameters outside the spinodal surface, see Fig. \ref{Fig:6}. The local stability was checked by numerically computing the Jacobian spectrum. 

It is worth noting that, in a one-dimensional model closely related to ours~\cite{Kourbane_Houssene_2018}, a hydrodynamic approach enabled the semi-analytical determination of the binodal line. More generally, Langevin-type simulations are commonly employed to locate the binodal line~\cite{Redner_2013, Fily2014, Su_2023, Hawthorne_2025}.

\section{Conclusion}
\label{sec4}

We used a master-equation formulation to study motility-induced phase separation in a lattice gas of persistent particles with hard-core exclusion. Bloch's theorem reduced the linearized dynamics around the homogeneous stationary state to a $z$-dimensional eigenvalue problem of tight-binding type. A perturbation expansion near $\vk = 0$ produced a closed-form spinodal surface that retains the same functional structure across the six Bravais lattices considered, with lattice geometry entering only through a single coefficient $\mathcal{A}$ that we computed exactly. Inhomogeneous stationary solutions inside the spinodal region become progressively smoother at the borders with increasing translational diffusion, in agreement with continuous-space Langevin studies of hard active disks~\cite{Hawthorne_2025}, and sustain rotational probability currents that follow directly from the absence of detailed balance and are computed here without coarse-graining.

The six Bravais lattices treated here are illustrative. The same construction could, in principle, be extended to other Bravais lattices, with or without a basis, and to dynamics with hopping beyond nearest neighbors.

\section*{Acknowledgements}

The authors would like to thank CNPq and CAPES for financial support and CCJDR for granting access to the HPC Cluster Coaraci, made available under FAPESP grants 2013/08293-7 and 2019/17874-0.

\bibliographystyle{ieeetr}  
\bibliography{ActiveLatticeModels}

\end{document}


\title{Supporting Information for Exact mean-field phase diagram for self-avoiding active particles in a lattice}
\author{Felipe Hawthorne}
\affiliation{Department of Physics, Federal University of Parana,
               R. Evaristo F. Ferreira da Costa, 81530-015, Curitiba, Brazil}
\affiliation{Interdisciplinary Center for Science, Technology, and Innovation (CICTI), Federal University of Parana, Av. Cel. Francisco H. dos Santos, 81530-000, Curitiba, Brazil}

\author{Cristiano F. Woellner}
\affiliation{Department of Physics, Federal University of Parana,
               R. Evaristo F. Ferreira da Costa, 81530-015, Curitiba, Brazil}
\affiliation{Interdisciplinary Center for Science, Technology, and Innovation (CICTI), Federal University of Parana, Av. Cel. Francisco H. dos Santos, 81530-000, Curitiba, Brazil}

\author{Jos\'e A. Freire}
\email[]{jfreire@fisica.ufpr.br}
\affiliation{Department of Physics, Federal University of Parana,
               R. Evaristo F. Ferreira da Costa, 81530-015, Curitiba, Brazil}
\maketitle

\section{The full eigenvalue spectrum of $\mJ$ in the square lattice}

\begin{figure}[h!]
\centering
{\includegraphics[width=\textwidth]{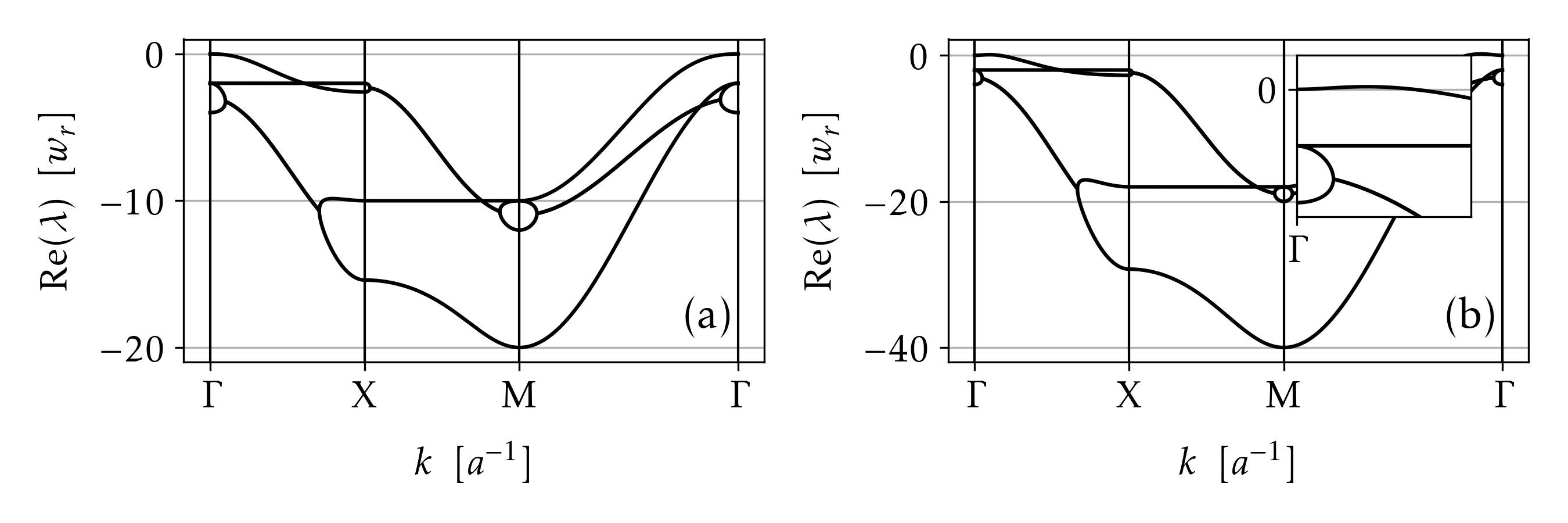}}
\caption{Real part of the fluctuation spectrum along high symmetry lines of the square lattice 1BZ, see Fig. \ref{fig:1}. (a) $w_t=0$, $w_a=10\,w_r$, and $\phi=0.6$. All eigenvalues have non-positive real parts. The homogeneous solution is stable. (b) $w_t=0$, $w_a=20\,w_r$, and $\phi=0.6$. The curvature of the uppermost ``band" becomes positive at the $\Gamma$ ($\vk=0$) point (better viewed in the inset). Some eigenvalues have positive real part. The homogeneous solution is unstable against long wavelength fluctuations.}\label{fig:2}
\end{figure}

\begin{figure}[h!]
\centering
{\includegraphics[width=1.5in]{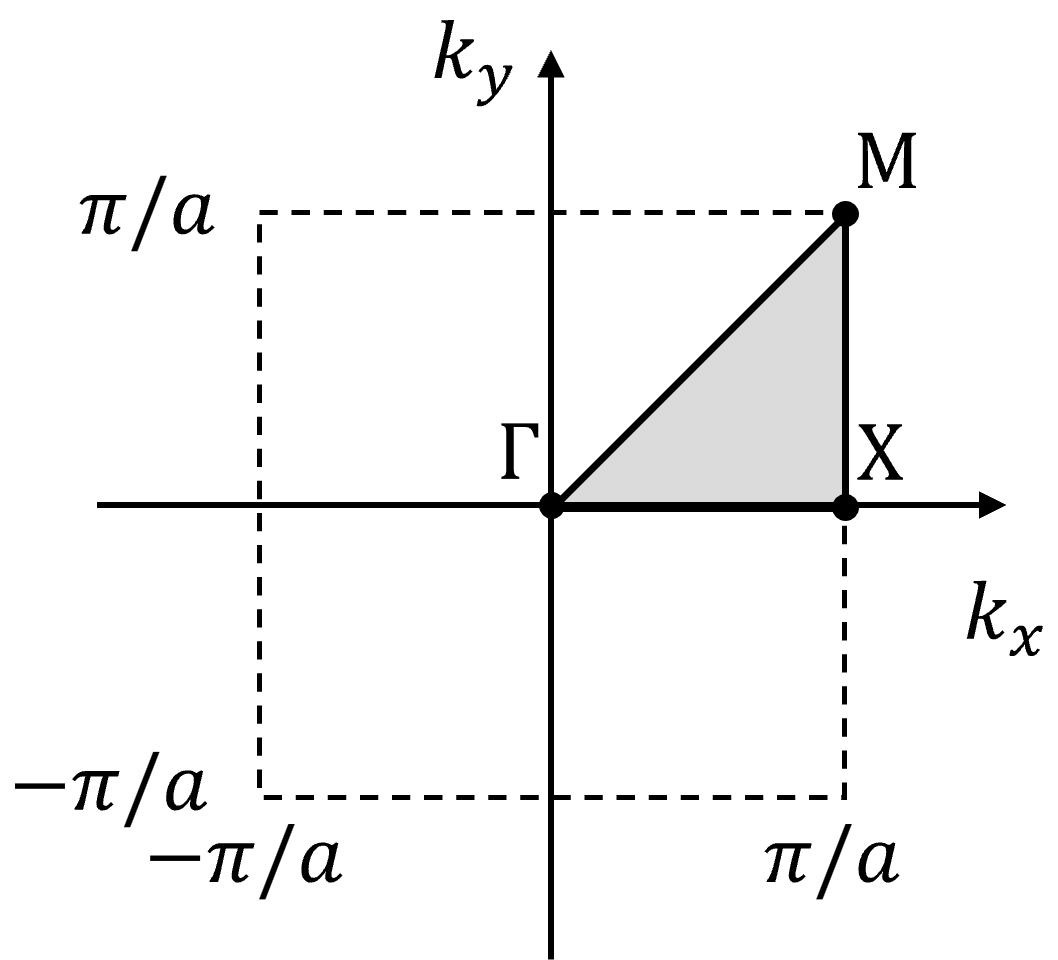}}
\caption{The first Brillouin zone of the square lattice and the high symmetry lines used in Fig. \ref{fig:2}.} \label{fig:1}
\end{figure}

Figure \ref{fig:2} illustrates the fluctuation spectrum in the case of a square lattice, whose first Brillouin zone (1BZ) is displayed in Fig. \ref{fig:1}. Two representative parameter sets $\{w_a, w_t, \phi\}$ ($w_a$ and $w_t$ expressed in units of $w_r$) were used: one corresponding to a stable homogeneous state, panel (a), and the other to an unstable one, panel (b). The last panel demonstrates that the homogeneous state becomes unstable with respect to long-wavelength fluctuations, as indicated by the emergence of positive eigenvalues in the vicinity of the $\vk=0$ point.

\section{The eigenvalues of $\mJ$ in the vicinity of $\vk=0$}
Here we derive the expression of $A$ in $\lambda_0=A \vk^2 +{\cal{O}}(k^3)$, as a function of the model parameters for six different Bravais lattices.

It is practical to use $w_r$ as unit of frequency and the lattice parameter, $a\equiv |\va_s|$, as the unit of length.
We start by repeating the relevant expressions from the main text ($d$ is the spatial dimension):

\begin{equation}
\begin{aligned}
    \mj_1 \equiv & \, w_a \, \sum_s i(\vk\cdot \va_s)\, \mA(s) \\
    \mj_2 \equiv & -  \frac{w_t z \vk^2}{2d} \, \mD -  \,
        \frac{w_a}{2} \sum_s (\vk\cdot \va_s)^2\, \mA(s)
\end{aligned}
\end{equation}
\begin{equation}
    \lambda_{0} = \vv_0^T \cdot \mj_1 \cdot \vv_0 + \vv_0^T \cdot \mj_2 \cdot \vv_0 - \sum_{n\ne 0} \frac{(\vv_0^T \cdot \mj_1 \cdot \vv_n)  (\vv_n^T \cdot \mj_1 \cdot \vv_0)}{\lambda_n}+\mathcal{O}(k^3) \label{PertTheory}
\end{equation}

The zero eigenvector of the $\mR$ matrix is $\vv_0= z^{-1/2}[1,1,...,1]^T$ in all lattices. From the general expressions for $\mD$ and $\mA(s)$ in the main text, we obtain:
\begin{equation}
    \vv_0^T \cdot \mA(s) \cdot \vv_0 = z^{-1},  \qquad 
    \vv_0^T \cdot \mD \cdot \vv_0 = 1
\end{equation}
The first two matrix elements in eq. \eqref{PertTheory} are then obtained:
\begin{equation}\label{J2}
     \vv_0^T \cdot \mJ_1 \cdot \vv_0 = 0, \qquad  \vv_0^T \cdot \mJ_2 \cdot \vv_0 = - \Big( \frac{w_t z + w_a}{2d} \Big) \vk^2,
\end{equation}
where we have used 
$$\sum_s \va_s \va_s=\Big(\frac{z}{d}\Big)\,\mI.$$

Regarding the third matrix element in \eqref{PertTheory}, we start with two results that directly follow from the general expression of the $\mA(s)$ matrix in the main text:
\begin{equation}
    \vv_0^T\cdot \mJ_1 = -\frac{iw_a(1-\phi)}{\sqrt{z}} \vu_{\vk}^T
\end{equation}
\begin{equation}
    \mJ_1 \cdot \vv_0= -\frac{iw_a(1-2\phi)}{\sqrt{z}} \vu_{\vk}
\end{equation}
with the definition ($k_i\equiv \vk\cdot \va_i$)
\begin{equation}
    \vu_\vk \equiv [k_1,k_2,...,k_z]^T.
\end{equation}
 
We then obtain:
\begin{equation}
    - \sum_{n\ne 0} \frac{(\vv_0^T \cdot \mJ_1 \cdot \vv_n)  (\vv_n^T \cdot \mJ_1 \cdot \vv_0)}{\lambda_n} = \bigg[\frac{w_a^2(1-\phi)(1-2\phi)}{z}\bigg]\,\vu_\vk^T \cdot \bigg( \sum_{n\ne 0} \frac{\vv_n \vv_n^T}{\lambda_n} \bigg) \cdot \vu_\vk
\end{equation}

We can write the matrix element in the RHS in a more practical form, exploiting the fact that $\vv_0^T \cdot \vu_\vk=0$:
$$
\vu_\vk^T \cdot \bigg( \sum_{n\ne 0} \frac{\vv_n \vv_n^T}{\lambda_n} + \vv_0 \vv_0^T\bigg) \cdot \vu_\vk= \vu_\vk^T \cdot (\mR+1)^{-1} \cdot \vu_\vk={\cal{A}}\, \vk^2,
$$
where ${\cal{A}}$ is a constant specific to each Bravais lattice (it does not depend on any of the model parameters).

The final result for the quadratic form associated with the zero eigenvalue is:
\begin{equation*}
\lambda_0 = \Bigg[ -\bigg( \frac{w_t z + w_a}{2d}\bigg) + \frac{w_a^2(1-\phi)(1-2\phi)}{z}\,{\cal{A}}\Bigg]\,\vk^2\qquad (\vk \sim 0)
\end{equation*}

One can show that, if the steric interaction is turned off, the above expression becomes, 
$$
(\lambda_0)_{\rm non~ inter.} = \Bigg[ -\bigg( \frac{w_t z + w_a}{2d}\bigg) + \frac{w_a^2}{z}\,{\cal{A}}\Bigg]\,\vk^2\qquad (\vk \sim 0).
$$
Since ${\cal{A}}<0$ for all Bravais lattices (see below), this implies that the homogeneous solution is unconditionally stable in the absence of interaction, as could be anticipated. No clustering without interaction. 

We obtain the constant ${\cal{A}}$ for six different Bravais lattices below.

\section{Calculation of the quadratic form $\vu_\vk^T \cdot (\mR+1)^{-1} \cdot \vu_\vk={\cal{A}}\, \vk^2$ for six different Bravais lattices}
Recall that $z$ is the lattice coordination number, $\va_s$ is the lattice vector in the coordinate direction $s=\{1,\ldots,z\}$, and $n_z$ is the number of directions adjacent to any given one (using a minimum angle criterion)


\subsection{Linear Lattice ($d=1, \, z=2, \, n_z=1)$}

\begin{figure}[h!]
\centering
{\includegraphics[width=1.5 in]{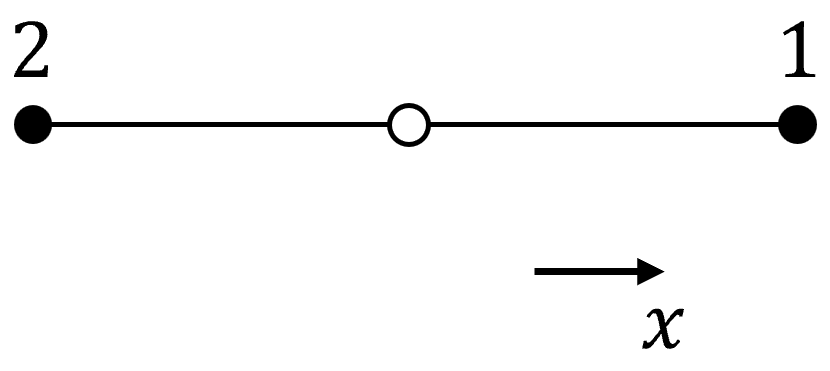}}
\caption{A particular labeling convention for the nearest neighbors of the white site in the linear lattice.}
\end{figure}

Using the labeling convention of the figure:
\begin{equation}
\begin{split}
\va_1 =&\, \hat{x} \\
\va_2 =&\, -\hat{x} 
\end{split}
\end{equation}

$$\vu_\vk =[k_x,-k_x]^T $$

\[
\mR = 
\begin{bmatrix}
-1 & 1   \\
1 & -1 
\end{bmatrix}
\]

$$\vu_\vk^T \cdot (\mR + 1)^{-1} \cdot \vu_\vk = -\vk^2 \to {\cal{A}}_{lin}=-1$$

$$
(\lambda_0)_{lin} = \Bigg[ -\bigg( \frac{2w_t + w_a}{2}\bigg) - \frac{w_a^2(1-\phi)(1-2\phi)}{2}\Bigg]\,\vk^2\qquad (\vk \sim 0)
$$

\subsection{Square Lattice ($d=2, \, z=4, \, n_z=2)$}

\begin{figure}[h!]
\centering
{\includegraphics[width=1.5 in]{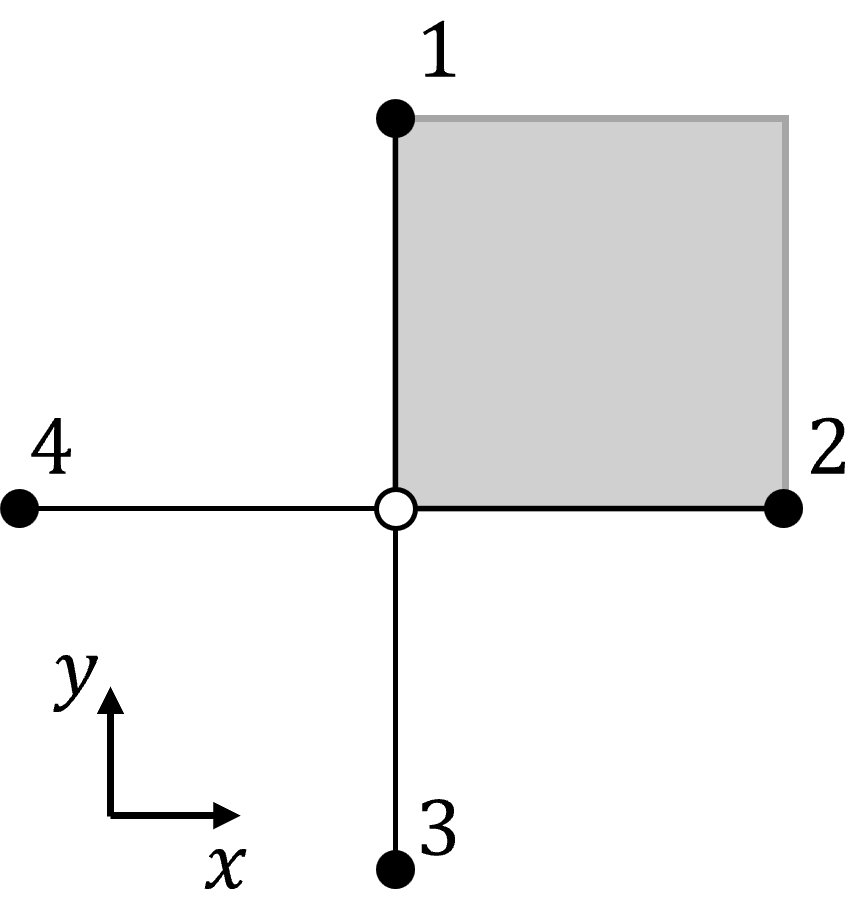}}
\caption{A particular labeling convention for the nearest neighbors of the white site in the square lattice.}\label{fig:square lattice}
\end{figure}

Using the labeling convention of the figure:
\begin{equation}
\begin{split}
\va_1 =&\, (0,1) \\
\va_2 =&\, (1,0) \\
\va_3 =&\, (0,\bo) \\
\va_4 =&\, (\bo,0) 
\end{split}
\end{equation}

According to the minimum angle adjacency criterion, the director $\va_1$ is adjacent to $\va_2$ and $\va_4$, etc. $(n_z=2)$.

$$\vu_\vk =[k_y,k_x, -k_y,-k_x]^T $$

\[
\mR = 
\begin{bmatrix}
-2 & 1 & \cdot & 1  \\
1 & -2 & 1 & \cdot   \\
\cdot & 1 & -2 & 1   \\
1 & \cdot & 1 & -2 
\end{bmatrix}
\]

$$\vu_\vk^T \cdot (\mR + 1)^{-1} \cdot \vu_\vk = -\vk^2 \to {\cal{A}}_{sq}=-1$$

$$
(\lambda_0)_{sq} = \Bigg[ -\bigg( \frac{4w_t + w_a}{4}\bigg) - \frac{w_a^2(1-\phi)(1-2\phi)}{4}\Bigg]\,\vk^2\qquad (\vk \sim 0)
$$

\subsection{Hexagonal Lattice $(d=2,\,z=6,\,n_z=2)$}
\begin{figure}[h!]
\centering
{\includegraphics[width=1.5 in]{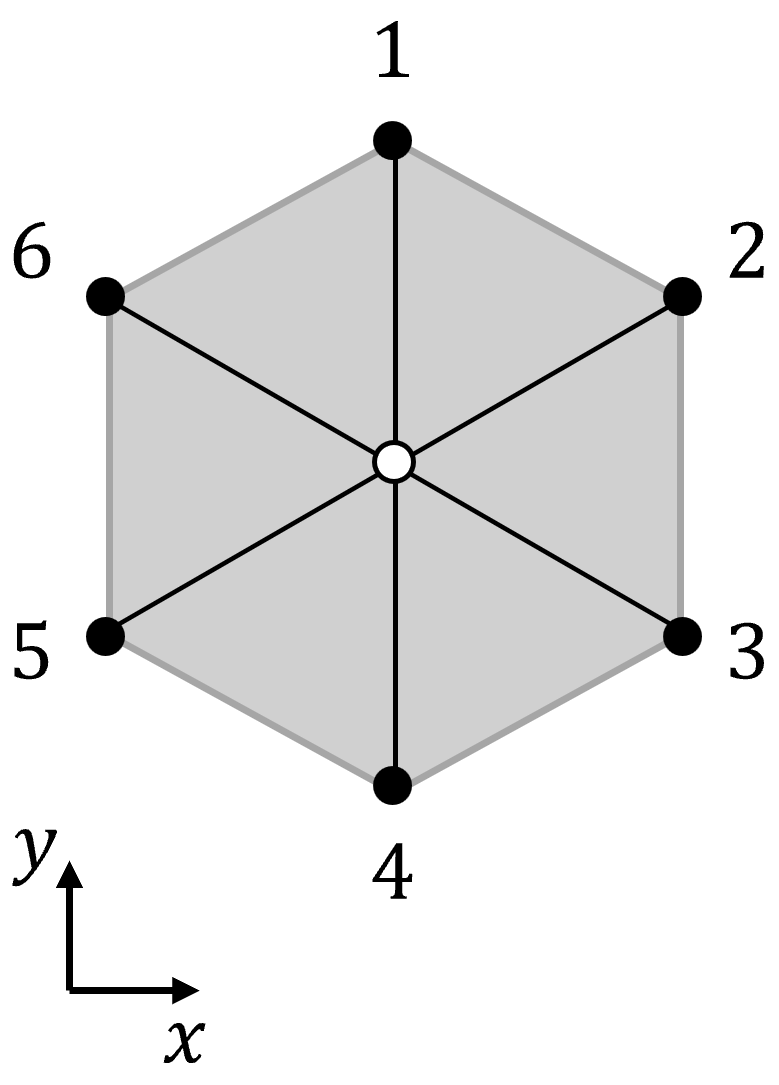}}
\caption{A particular labeling convention for the nearest neighbors of the white site in the hexagonal lattice.}
\end{figure}

Using the labeling convention of the figure:
\begin{equation}
\begin{split}
\va_1 =&\, (0,1) \\
\va_2 =&\, (\tfrac{\sqrt{3}}{2},\tfrac{1}{2}) \\
\va_3 =&\, (\tfrac{\sqrt{3}}{2},-\tfrac{1}{2}) \\
\va_4 =&\, (0,-1) \\
\va_5 =&\, (\tfrac{-\sqrt{3}}{2},-\tfrac{1}{2}) \\
\va_6 =&\, (\tfrac{-\sqrt{3}}{2},\tfrac{1}{2}) 
\end{split}
\end{equation}

According to the minimum angle adjacency criterion, the director $\va_1$ is adjacent to $\va_2$ and $\va_6$, etc. $(n_z=2)$.

$$\vu_\vk =[k_y,\tfrac{\sqrt{3}}{2}k_x+\tfrac{1}{2}k_y,\ldots, -\tfrac{\sqrt{3}}{2}k_x+\tfrac{1}{2}k_y]^T $$

\[
\mR = 
\begin{bmatrix}
-2 & 1 & \cdot & \cdot & \cdot & 1  \\
1 & -2 & 1 & \cdot & \cdot & \cdot  \\
\cdot & 1 & -2 & 1 & \cdot & \cdot  \\
\cdot & \cdot & 1 & -2 & 1 & \cdot  \\
\cdot & \cdot & \cdot & 1 & -2 & 1  \\
1 & \cdot & \cdot & \cdot & 1 & -2 
\end{bmatrix}
\]

$$\vv_\vk^T \cdot (\mR + 1)^{-1} \cdot \vv_\vk = - 3\vk^2 \to {\cal{A}}_{hex}=-3$$

$$
(\lambda_0)_{hex} = \Bigg[ -\bigg( \frac{6w_t + w_a}{4}\bigg) - \frac{w_a^2(1-\phi)(1-2\phi)}{2}\Bigg]\,\vk^2\qquad (\vk \sim 0)
$$

\subsection{Simple Cubic Lattice $(d=3,\,z=6,\,n_z=4)$}

\begin{figure}[h!]
\centering
{\includegraphics[width=2.5in]{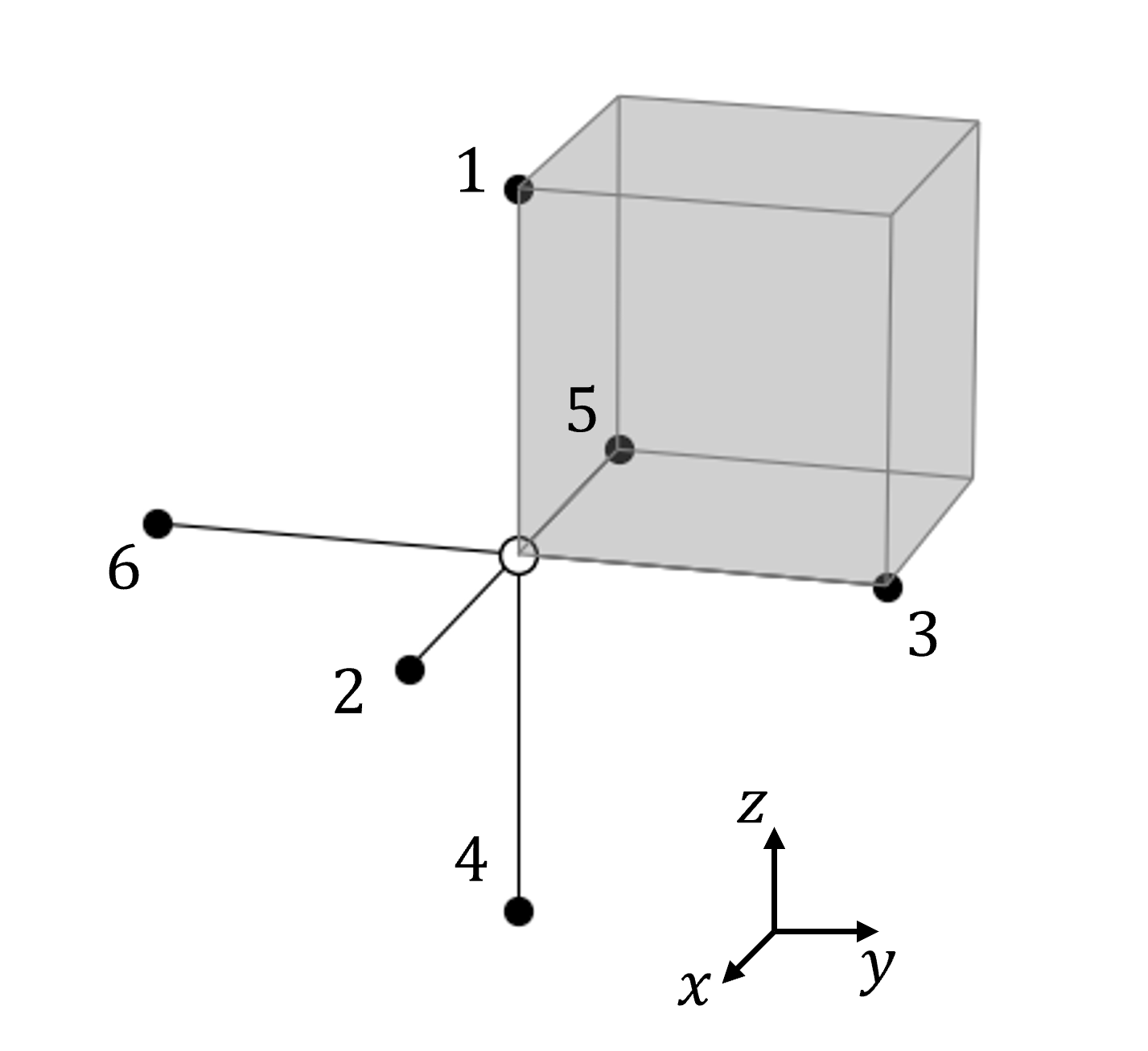}}
\caption{A particular labeling convention for the nearest neighbors of the white site in the simple cubic lattice.}
\end{figure}

Using the labeling convention of the figure:
\begin{equation}
\begin{split}
\va_1 =&\, (0,0,1) \\
\va_2 =&\, (1,0,0) \\
\va_3 =&\, (0,1,0) \\
\va_4 =&\, (0,0,\bo) \\
\va_5 =&\, (\bo,0,0) \\
\va_6 =&\, (0,\bo,0) 
\end{split}
\end{equation}

According to the minimum angle adjacency criterion, the director $\va_1$ is adjacent to $\va_2$, $\va_3$, $\va_5$, and $\va_6$, etc. $(n_z=4)$.

With this choice of labeling, 
$$\vu_\vk = [k_z, k_x, \ldots, -k_y]^T $$

\[
\mR = 
\begin{bmatrix}
-4 & 1 & 1 & \cdot & 1 & 1  \\
1 & -4 & 1 & 1 & \cdot & 1  \\
1 & 1 & -4 & 1 & 1 & \cdot  \\
\cdot & 1 & 1 & -4 & 1 & 1  \\
1 & \cdot & 1 & 1 & -4 & 1  \\
1 & 1 & \cdot & 1 & 1 & -4 
\end{bmatrix}
\]

$$\vu_\vk^T \cdot (\mR + 1)^{-1} \cdot \vu_\vk = -\frac{\vk^2}{2} \to {\cal{A}}_{sc}=-\frac{1}{2}$$

$$
(\lambda_0)_{sc} = \Bigg[ -\bigg( \frac{6w_t + w_a}{6}\bigg) - \frac{w_a^2(1-\phi)(1-2\phi)}{12}\Bigg]\,\vk^2\qquad (\vk \sim 0)
$$

\subsection{Body Centered Cubic Lattice $(d=3,\,z=8,\,n_z=3)$}

\begin{figure}[h!]
\centering
{\includegraphics[width=3in]{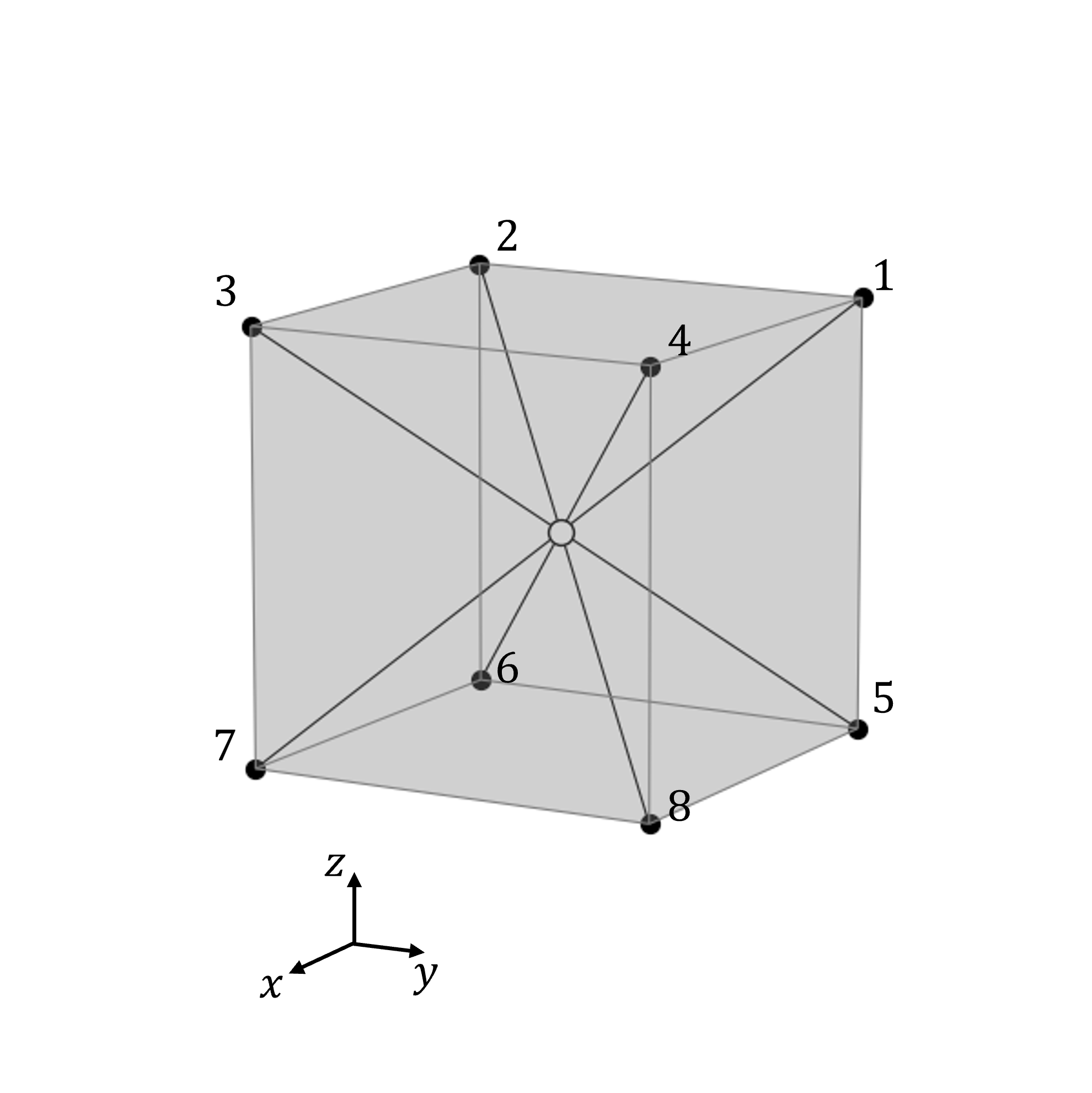}}
\caption{A particular labeling convention for the nearest neighbors of the white site in the body centered cubic lattice.}
\end{figure}

Using the labeling convention of the figure:
\begin{equation}
\begin{split}
\va_1 =&\, \tfrac{1}{\sqrt{3}}(\bo,1,1) \\
\va_2 =&\, \tfrac{1}{\sqrt{3}}(\bo,\bo,1) \\
\va_3 =&\, \tfrac{1}{\sqrt{3}}(1,\bo,1) \\
\va_4 =&\, \tfrac{1}{\sqrt{3}}(1,1,1) \\
\va_5 =&\, \tfrac{1}{\sqrt{3}}(\bo,1,\bo) \\
\va_6 =&\, \tfrac{1}{\sqrt{3}}(\bo,\bo,\bo) \\
\va_7 =&\, \tfrac{1}{\sqrt{3}}(1,\bo,\bo) \\
\va_8 =&\, \tfrac{1}{\sqrt{3}}(1,1,\bo) \\
\end{split}
\end{equation}

According to the minimum angle adjacency criterion, the director $\va_1$ is adjacent to $\va_2$, $\va_4$, and $\va_5$, etc. $(n_z=3)$.

$$\vu_\vk = [\tfrac{1}{\sqrt{3}}(-k_x+k_y+k_z), \tfrac{1}{\sqrt{3}}(-k_x-k_y+k_z), \ldots, \tfrac{1}{\sqrt{3}}(k_x+k_y-k_z)]^T $$

\[
\mR = 
\begin{bmatrix}
-3 & 1 & \cdot & 1 & 1 & \cdot & \cdot & \cdot \\
1 & -3 & 1 & \cdot & \cdot & 1 & \cdot & \cdot \\
\cdot & 1 & -3 & 1 & \cdot & \cdot & 1 & \cdot \\
1 & \cdot & 1 & -3 & \cdot & \cdot & \cdot & 1 \\
1 & \cdot & \cdot & \cdot & -3 & 1 & \cdot & 1 \\
\cdot & 1 & \cdot & \cdot & 1 & -3 & 1 & \cdot \\
\cdot & \cdot & 1 & \cdot & \cdot & 1 & -3 & 1 \\
\cdot & \cdot & \cdot & 1 & 1 & \cdot & 1 & -3
\end{bmatrix}
\]

$$\vu_\vk^T \cdot (\mR + 1)^{-1} \cdot \vu_\vk = - \frac{4\vk^2}{3} \to {\cal{A}}_{bcc}=-\frac{4}{3}$$

$$
(\lambda_0)_{bcc} = \Bigg[ -\bigg( \frac{8w_t + w_a}{6}\bigg) - \frac{w_a^2(1-\phi)(1-2\phi)}{6}\Bigg]\,\vk^2\qquad (\vk \sim 0)
$$

\subsection{Face Centered Cubic Lattice $(d=3,\,z=12,\,n_z=4)$}

\begin{figure}[h!]
\centering
{\includegraphics[width=2.5in]{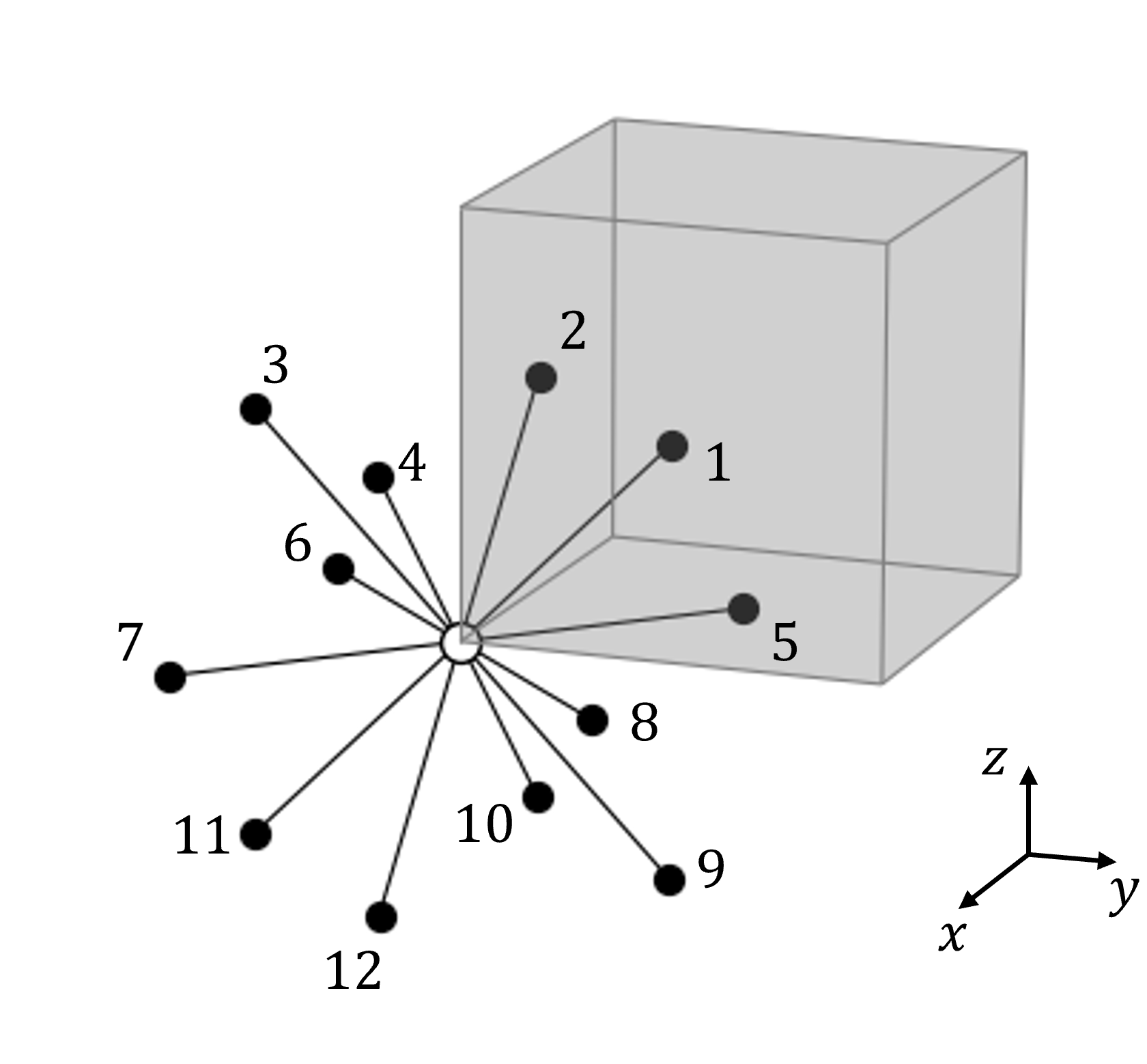}}
\caption{A particular labeling convention for the nearest neighbors of the white site in the face centered cubic lattice.}
\end{figure}

Using the labeling convention of the figure:
\begin{equation}
\begin{split}
\va_1 =&\, \tfrac{1}{\sqrt{2}}(0,1,1) \\
\va_2 =&\, \tfrac{1}{\sqrt{2}}(\bo,0,1) \\
\va_3 =&\, \tfrac{1}{\sqrt{2}}(0,\bo,1) \\
\va_4 =&\, \tfrac{1}{\sqrt{2}}(1,0,1) \\
\va_5 =&\, \tfrac{1}{\sqrt{2}}(\bo,1,0) \\
\va_6 =&\, \tfrac{1}{\sqrt{2}}(\bo,\bo,0) \\
\va_7 =&\, \tfrac{1}{\sqrt{2}}(1,\bo,0) \\
\va_8 =&\, \tfrac{1}{\sqrt{2}}(1,1,0) \\
\va_9 =&\, \tfrac{1}{\sqrt{2}}(0,1,\bo) \\
\va_{10} =&\, \tfrac{1}{\sqrt{2}}(\bo,0,\bo) \\
\va_{11} =&\, \tfrac{1}{\sqrt{2}}(0,\bo,\bo) \\
\va_{12} =&\, \tfrac{1}{\sqrt{2}}(1,0,\bo)
\end{split}
\end{equation}

According to the minimum angle adjacency criterion, the director $\va_1$ is adjacent to $\va_2$, $\va_4$, $\va_5$, and $\va_8$, etc. $(n_z=4)$.

$$\vu_\vk = [\tfrac{1}{\sqrt{2}}(k_y+k_z), \tfrac{1}{\sqrt{2}}(-k_x+k_z), \ldots, \tfrac{1}{\sqrt{2}}(k_x-k_z)]^T $$

\[
\mR = 
\begin{bmatrix}
-4 & 1 & \cdot & 1 & 1 & \cdot & \cdot & 1 & \cdot & \cdot & \cdot & \cdot \\
1 & -4 & 1 & \cdot & 1 & 1 & \cdot & \cdot & \cdot & \cdot & \cdot & \cdot \\
\cdot & 1 & -4 & 1 & \cdot & 1 & 1 & \cdot & \cdot & \cdot & \cdot & \cdot \\
1 & \cdot & 1 & -4 & \cdot & \cdot & 1 & 1 & \cdot & \cdot & \cdot & \cdot \\
1 & 1 & \cdot & \cdot & -4 & \cdot & \cdot & \cdot & 1 & 1 & \cdot & \cdot \\
\cdot & 1 & 1 & \cdot & \cdot & -4 & \cdot & \cdot & \cdot & 1 & 1 & \cdot \\
\cdot & \cdot & 1 & 1 & \cdot & \cdot & -4 & \cdot & \cdot & \cdot & 1 & 1 \\
1 & \cdot & \cdot & 1 & \cdot & \cdot & \cdot & -4 & 1 & \cdot & \cdot & 1 \\
\cdot & \cdot & \cdot & \cdot & 1 & \cdot & \cdot & 1 & -4 & 1 & \cdot & 1 \\
\cdot & \cdot & \cdot & \cdot & 1 & 1 & \cdot & \cdot & 1 & -4 & 1 & \cdot \\
\cdot & \cdot & \cdot & \cdot & \cdot & 1 & 1 & \cdot & \cdot & 1 & -4 & 1 \\
\cdot & \cdot & \cdot & \cdot & \cdot & \cdot & 1 & 1 & 1 & \cdot & 1 & -4
\end{bmatrix}
\]

$$\vu_\vk^T \cdot (\mR + 1)^{-1} \cdot \vu_\vk = - 2\vk^2 \to {\cal{A}}_{fcc}=-2$$

$$
(\lambda_0)_{fcc} = \Bigg[ -\bigg( \frac{12w_t + w_a}{6}\bigg) - \frac{w_a^2(1-\phi)(1-2\phi)}{6}\Bigg]\,\vk^2\qquad (\vk \sim 0)
$$

\section{The general Jacobian expression}
In the main text we computed the Jacobian for the homogeneous solution. Here we compute its general expression. We obtain that, owing to the active motion and the steric interaction, $\mJ$ is not a symmetrical matrix, which points to the fact that the RHS of the mean-field master equation \emph{is not} a gradient, i.e., the stationary solutions \emph{cannot} be viewed as minima of some free-energy function. 

We start by reproducing the mean-field master equation, adding a parameter $\epsilon$ to keep track of the steric interaction in the final expression.
\begin{equation}
\begin{aligned}
  \frac{dp_{\vr s}}{dt} = & ~ w_a \, \big[ p_{\vr-\va_s s} \, (1 - \epsilon\, p_{\vr\Sigma}) - p_{\vr s} \, (1 - \epsilon \,p_{\vr+\va_s\Sigma}) \big] \\
+ & ~ w_t \,\sum_{\vr'\sim \,\vr} \big[ p_{\vr' s} \, (1 - \epsilon \,p_{\vr\Sigma})
-  p_{\vr s} \, (1 - \epsilon\, p_{\vr' \Sigma} ) \big] \\
+ & ~ w_r \, \sum_{s' \sim \,s} (p_{\vr s'} - p_{\vr s}).
\end{aligned}
\end{equation}

The Jacobian matrix is defined as
$$\mJ_{\vr s,\tilde \vr \tilde s}\equiv \frac{\partial (RHS)}{\partial p_{\tilde \vr \tilde s}}$$

We compute the contribution of each term in separate below.

The active motion term,
\begin{align*}
    \mJ_{\vr s,\tilde \vr \tilde s}^{(a)} = 
    \omega_a \Big[& \delta_{\vr, \tilde \vr + \va_s} \,\delta_{s,\tilde s} \,(1-\epsilon\, p_{\vr\Sigma}) - \epsilon\,\delta_{\vr,\tilde \vr} \, p_{\vr-\va_s s} \\
    & - \delta_{\vr,\tilde \vr}\, \delta_{s,\tilde s} (1-\epsilon\,p_{\vr+\va_s \Sigma})
    + \epsilon\,\delta_{\vr, \tilde \vr - \va_s}\, p_{\vr s}
    \Big],
\end{align*}
\emph{is not} symmetrical, even in the absence of steric interaction $(\epsilon=0)$.

The translational diffusion term,
\begin{align*}
    \mJ_{\vr s,\tilde \vr \tilde s}^{(t)} = 
    \omega_t \Big[& \delta_{\vr\sim \,\tilde \vr} \,\delta_{s,\tilde s} \,(1-\epsilon\, p_{\vr\Sigma}) - \epsilon\,\delta_{\vr,\tilde \vr} \sum_{\vr'\sim \,\vr} p_{\vr's} \\
    &- \delta_{\vr,\tilde \vr}\, \delta_{s,\tilde s} \sum_{\vr'\sim \vr} (1-\epsilon\,p_{\vr'\Sigma})
    + \epsilon\,\delta_{\vr\sim \tilde \vr}\, p_{\vr s}
    \Big]  ,
\end{align*}
is symmetrical only in the absence of steric interaction $(\epsilon=0)$.

The rotational diffusion term,
\begin{equation*}
    \mJ_{\vr s,\tilde \vr \tilde s}^{(r)} = \omega_r \, \delta_{\vr,\tilde \vr}\, [-n_z \delta_{s,\tilde s} + \delta_{s\sim \tilde s}],
\end{equation*}
is symmetrical. 

In conclusion, the presence of activity and steric interactions is what makes the mean-field master equation non-gradient, and implies that its stationary states cannot be viewed as extrema of a free energy.

Substituting $p_{\vr s}=\phi/z$ reproduces the Jacobian presented in the main text.

\section{The probability current density}
The ME can be viewed as a discrete probability continuity equation.
We identify as the probability current between the sites $\vr$ and $\vr+\va_s$:
\begin{equation}
\begin{aligned}
    \vJ_{\vr+\tfrac{\va_s}{2}} = \frac{w_a}{{\cal{A}}} & \big[ p_{\vr,s}(1-p_{\vr+\va_s,\Sigma})-p_{\vr+\va_s,-s}(1-p_{\vr,\Sigma}) \big]\,\hat{\va}_s  \\
    + \frac{w_t}{{\cal{A}}} & \big[ p_{\vr,\Sigma}-p_{\vr+\va_s,\Sigma} \big]\,\hat{\va}_s
\end{aligned}
\end{equation}
where ${\cal{A}}$ is the lenght/area of the unit cell face. Note the translational diffusion current is a discrete version of Fick's current.

The ME implies the following continuity equation for the total occupation of site $\vr$:
\begin{equation}
    \frac{dp_{\vr,\Sigma}}{dt} \equiv \sum_s \vJ_{\vr+\tfrac{\va_s}{2}} \cdot ({\cal{A}} \,\hat{\va}_s)
\end{equation}

We define the probability current at $\vr$ as the sum of the averages between opposing cell faces:
\begin{equation}
    \vJ_\vr = \frac{1}{2} \sum_s \vJ_{\vr+\tfrac{\va_s}{2}}.
\end{equation}

\section{About $\Delta S(t)$}

\begin{figure}[h!]
\centering
{\includegraphics[width=0.6\textwidth]{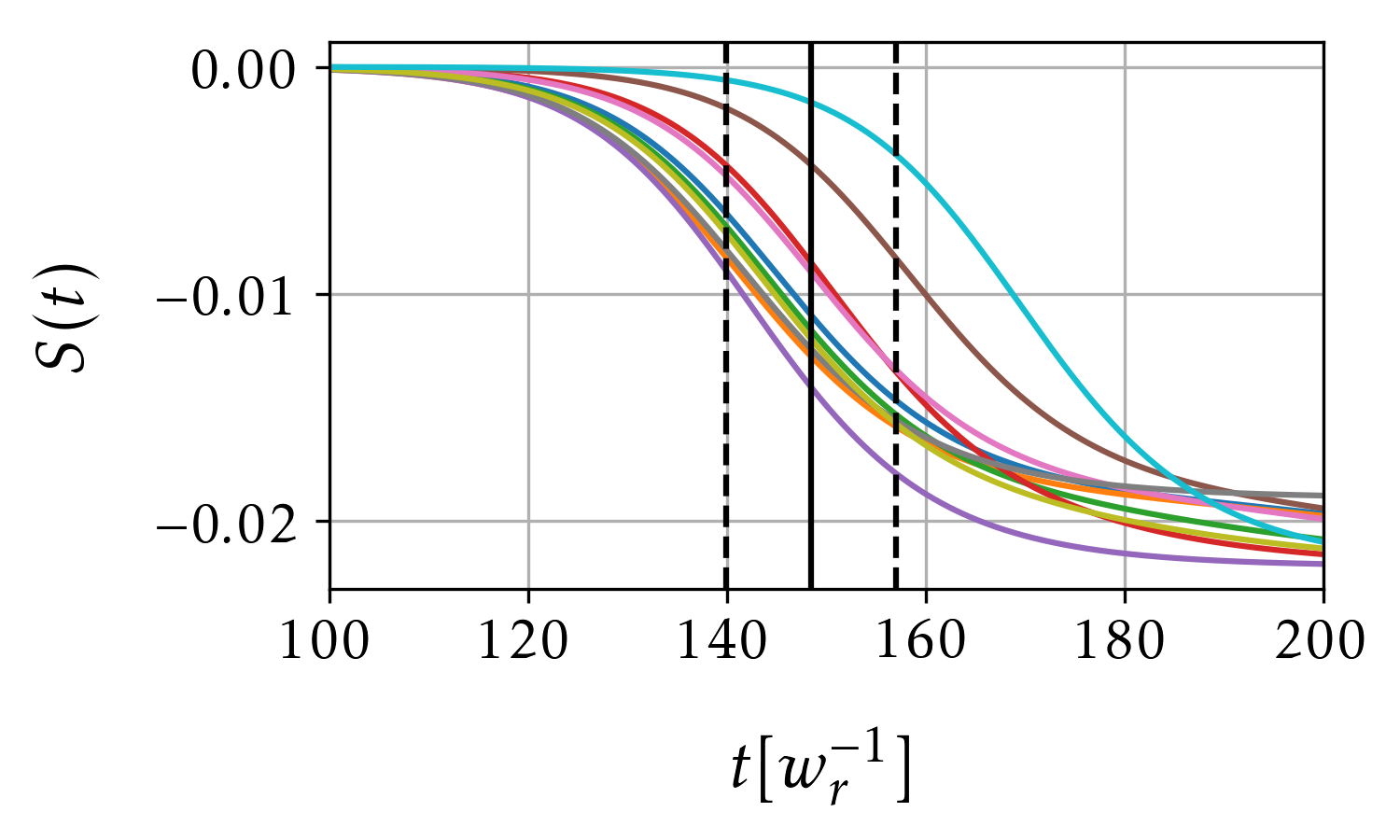}}
\caption{Typical time evolution of $\Delta S(t)$, the entropy per particle of the current PDF $\vp(t)$, relative to the homogeneous state PDF, for ten different random fluctuations about the homogeneous state. We associated the escape time with the first inflection point. The solid line is the average and the dashed lines are one standard deviation from the average. Model parameters: $w_a=49.58\,w_r$, $w_t=50\,w_r$, and $\phi=0.7458$, on an $80\times 80$ square lattice.}\label{fig:DS(t)}
\end{figure}

\begin{figure}[h!]
\centering
{\includegraphics[width=0.8\textwidth]{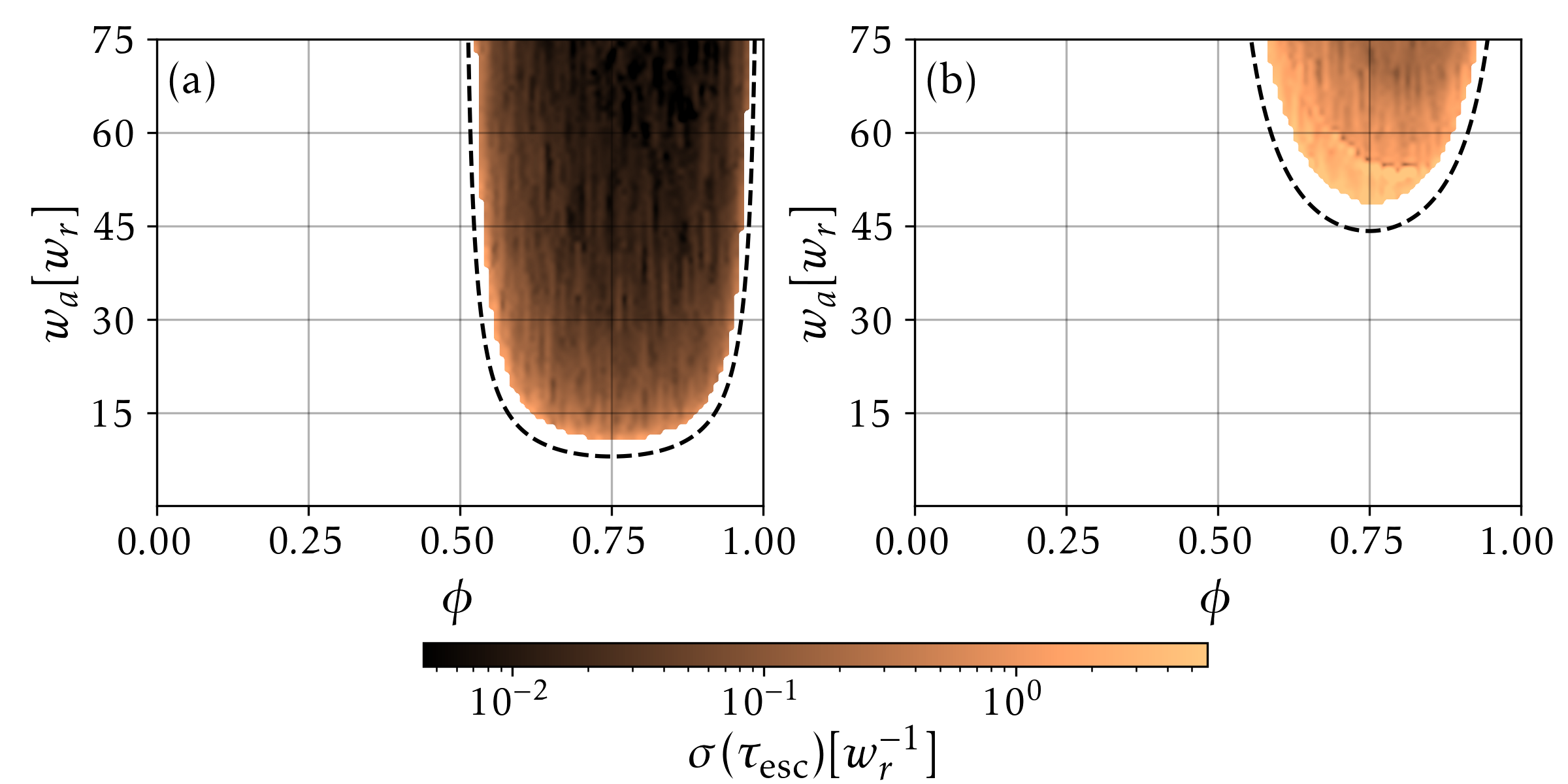}}
\caption{Standard deviation of the escape time numerical measurement over ten different initial conditions in a $80\times 80$ square lattice. The dashed lines are the infinite square lattice spinodal lines at the corresponding value of $w_t$. Panel (a) corresponds to $w_t=0$, and panel (b) to $w_t=50$.}\label{fig:escapestd}
\end{figure}

In the main text, we measured the escape time from the locally unstable homogeneous state as the first inflection point of $\Delta S(t)$, the entropy per particle of the current PDF $\vp(t)$ relative to the homogeneous state PDF, averaged over ten realizations of the random initial fluctuation about the homogeneous state. Figure \ref{fig:DS(t)} displays a representative set of such $\Delta S(t)$ trajectories, used to extract a single point of the $\tau_{esc}$ map shown in Fig. 5 of the main text. The corresponding standard deviation of the escape time across the spinodal region is reported in Fig. \ref{fig:escapestd}.

